# Wiley 5G Ref

# Article Template

### Article title:  **Beamforming and precoding techniques**

First author: Full name and affiliation; plus email address if corresponding author
Donatella Darsena, University Parthenope of Napoli

Second author: Full name and affiliation; plus email address if corresponding author
Giacinto Gelli, University Federico II of Napoli

Third author: Full name and affiliation; plus email address if corresponding author
Francesco Verde, University Federico II of Napoli,  email: f.verde@unina.it

## Word Count
12000

## Abstract

Beamforming and precoding/combining are techniques aimed at processing multiantenna signals at the transmitter and/or at the receiver of a wireless communication system. While they have been routinely used to improve performance in current and previous generations of mobile communications systems, they are expected to play a more fundamental role in 5th Generation (5G) New Radio (NR) cellular systems, whose functionalities have been defined in the first phase of 3GPP 5G standardization process. Besides operating in traditional cellular sub-6 GHz frequency band, 5G NR has been natively designed also to work in the higher millimeter-wave (MMW) band. At lower frequencies, multiantenna techniques for 5G NR are mainly refinements of those originally designed for 4G Long Term Evolution (LTE). On the contrary, to cope with the peculiarities of MMW scenarios, such as the larger number of antenna elements, the more directional transmission, and the higher path loss values, new dynamic, user-specific, and computationally-efficient multiantenna solutions and procedures have been incorporated in 5G NR specifications. In particular, since multiantenna techniques for 5G NR generally need detailed channel state information (CSI), a complete redesign of the set of reference signals and procedures used for CSI acquisition and reporting was carried out. 5G NR is continuously evolving and new features will be added, while the existing ones will be enhanced in the second phase of 5G standardization, with particularly emphasis on beam management operations, reduction of CSI overhead,  unconventional transmission methods, robustness against spatial correlation among channels, and software-based reconfigurable antennas.

## Main text

### [A] Introduction

In December 2017, the 3rd Generation Partnership Project (3GPP) (3GPP homepage, 2019) issued the first release (Rel-15) of the 5th Generation (5G) New Radio (NR) technology, namely, the new Radio Access Network (RAN) architecture for 5G systems. Initially, Rel-15 contained specifications for *non-standalone* (NSA) 5G systems, where user-plane functions are supported by the new RAN, while control-plane functions are still demanded to the 4th Generation (4G) Long Term Evolution (LTE) access and core network. Later on (2018), Rel-15 incorporated also specifications for *standalone* (SA) 5G systems, where 5G RAN takes care of all the user- and control-plane functions and is connected to a native cloud-based 5G core network.

Rel-15 ("frozen" in 2018) is informally referred to as "5G Phase 1", whose focus is mainly on *enhanced mobile broadband* (eMBB) and *ultra-reliable and low-latency communication* (URLLC) use cases, whereas support of *massive machine type communication* (mMTC) will still rely on LTE-based technologies, such as *enhanced machine-type communication* (eMTC) and *narrowband Internet of Things* (NB-IoT). Native NR support for mMTC and device-to-device (D2D) communications, called *sidelink transmission* in 5G, will be addressed in later releases. Rel-16 (expected to be finalized by 2020) is informally known as "5G Phase 2". It will provide significant advances over Rel-15 in several fields, such as operation in unlicensed spectrum (NR-U), solutions for integrated access/backhaul (IAB), industrial IoT, vehicle-to-everything (V2x) connectivity, enhancements of URLLC services, and enhanced multiantenna signal processing.

Within Rel-15, NR is allowed to operate (3GPP TS 38.104, 2019) in two frequency ranges:

1) from 410 MHz to 7125 MHz (FR1), also referred to as *sub-6 GHz* band;

2) from 24 250 MHz to 52 600 MHz (FR2), also referred to as *mmWave* (MMW) *band*.

In comparison, LTE systems mainly work in a subset of FR1, with operating frequency generally below 3 GHz. 5G should not be considered as a synonymous of MMW, since the new standard provides high flexibility and support for a wide range of frequencies. However, operation at MMW frequencies will represent a distinctive feature of NR, assuring substantial benefits (wider bandwidth and advanced spatial reuse) but involving many new challenges (higher propagation losses, increased sensitivity to blockage, innovative radio hardware and antenna design, faster support of mobile operations), which require a significant technical breakthrough over 4G systems.

The radio interface of 5G NR (3GPP TS 38.201, 2017) encompasses Layer 1 (Physical, PHY), Layer 2 (Medium Access Control, MAC), and Layer 3 (Radio Resource Control, RRC). PHY specifications are in the 38.200 series documents, whereas MAC and RRC layers are covered in the 38.300 series documents (the most important 3GPP NR documents for the purposes of this article are listed in Table 1). The NR PHY is based on *Orthogonal Frequency-Division Multiplexing* (OFDM) with cyclic prefix (CP) (5GRef001), with optional support for Discrete Fourier Transform (DFT) precoded OFDM (DFT-s-OFDM) in the UL. Both *frequency-division duplexing* (FDD) and *time-division duplexing* (TDD) schemes are envisioned to support operation in paired (mainly in FR1) as well as unpaired (mainly in FR2) bandwidths.

*Multiantenna techniques* represent an important enabling technology in NR PHY, in order to satisfy the challenging requirements of ITU IMT-2020 (ITU-R M.2083, 2015), in particular support of data-rates at least of 100 Mb/s in wide areas, with peak values of 20 Gb/s in hotspots, as well as 3x spectral efficiency or 100x energy efficiency compared to LTE-Advanced networks.

Using multiple antennas at the transmitting (TX) and/or receiving (RX) side of a wireless link offers different advantages to the system designer. Multiple antennas can be used in *transmit/receive diversity* configurations, to improve communication reliability against channel adverse effects (small-scale fading). They can also be utilized to increase spectral efficiency (i.e., data throughput) by *spatial multiplexing* (SM), where multiple data-streams (called *layers* in the 3GPP terminology) are transmitted in parallel using the same time/frequency resource. Finally, multiple antennas can be used for *transmit/receive beamforming*, which is a convenient solution to improve radio coverage, perform spatial reuse, and reduce co-channel interference.

*Beamforming* is a term traditionally associated with array processing or smart antennas (Van Veen, 1998). More recently, multidimensional signal processing techniques for multiantenna systems have also been referred to as *precoding* at the TX and *decoding* (or *combining*) at the RX. Since there is no common consensus about the meaning of the terms "beamforming" and "precoding" in the technical literature (Björnson, 2017), we will use them interchangeably to denote any kind of spatial processing of signals in a multiantenna system.

It should be observed that, unlike diversity and beamforming techniques, which work also in *single-input multiple-output* (SIMO) or *multiple-input single-output* (MISO) configurations, SM requires installation of multiple antennas at both the receiver and the transmitter, implementing a *multiple-input multiple-output* (MIMO) system. In a MIMO system employing SM, multiple layers in a given time/frequency resource can be transmitted to a single user equipment (UE), which is called *single-user MIMO* (SU-MIMO) or, by adopting aggressive spatial reuse strategies, to multiple UEs, referred to as *multi-user MIMO* (MU-MIMO) (Gesbert, 2007).

A multiantenna system with $N_T$ TX and $N_R$ RX antennas can provide different performance gains (Paulraj, 2003):

- *array gain*: signal-to-noise ratio (SNR) improvement due to coherent combination at the RX and/or coherent precoding at the TX, whose maximum value is upper bounded by $N_T$ or $N_R$, in MISO/SIMO configurations, respectively, or by the product $N_T \cdot N_R$ in MIMO configurations;

- *diversity gain*: improvement of the slope of the bit-error-rate (BER) versus SNR curve, which for Rayleigh fading channels is upper bounded by $N_T \cdot N_R$;

- *multiplexing gain*: improvement in capacity/data-rate, which is upper bounded by $\min(N_T, N_R)$.

Several trade-offs exist when trying to achieve these gains in practical systems with a finite number of antennas, such as the trade-off between diversity and SM (Zheng, 2003) or the trade-off between beamforming and SM (Sanchez-Fernandez, 2007).

Multiantenna techniques are expected to play a significant role in NR at different operating frequencies for different reasons:

1) at lower frequencies (i.e., in FR1), where the available bandwidth is scarce, communications are *bandwidth-limited*. SM techniques will be used (similarly to LTE) to improve spectral efficiency, achieving moderate-to-high data-rates, especially in wide coverage scenarios;

2) at higher frequencies (i.e., in FR2), where larger portions of bandwidth are available, communications are *power-limited*, because of higher path losses and blockage phenomena. Beamforming techniques will be used to provide the necessary SNR gain and reduce co-channel interference, by using highly-directional beams, especially in local coverage scenarios.

Multiantenna techniques were already at the center of LTE RAN specifications, with support for diversity, beamforming, and SM, providing transmission of up to 4 layers in *downlink* (DL) and of a single layer in *uplink* (UL) in Rel-8 (later releases significantly extended these functionalities). Thus, it is not surprising that many multiantenna solutions for NR are natural evolutions of those employed in LTE, especially when working at lower FR1 frequencies. However, the adoption of multiantenna techniques at higher MMW frequencies typical of FR2 will be a disruptive and challenging technology in NR, for several reasons.

First, working with smaller wavelength will pave the way to diffuse implementation of *massive MIMO* (mMIMO) systems (Larsson, 2014), that is, systems employing a huge number of antennas at the BS, generally much larger than the number of transmitted data streams. The benefits of mMIMO have been assessed by a large body of technical literature (Björnson, 2016) and some testbeds have confirmed (Gozalvez, 2016) that such advantages can be obtained in practical cases. As the number of TX antennas increases, two important facts (Marzetta, 2010) have been theoretically predicted: (1) the cross-correlation between any two random channel realizations vanishes, so that separations of MU-MIMO transmissions can be implemented even with simple *linear precoding* (LP); (2) the small-scale fading of each channel is averaged out, making the channel behave as a deterministic one (*channel hardening*).

Practical mMIMO implementations rely on *two-dimensional* (2D) array structures, to accommodate a large number of antennas without significantly increasing the deployment space. 3GPP started standardization of 2D array structures with the name of *full-dimensional MIMO* (FD-MIMO) in LTE Rel-13 (also known as LTE-Advanced Pro) (Ji, 2017), which utilizes up to 16 antenna ports, extended to 32 antenna ports in LTE Rel-14. NR supports 32 antenna ports from its first release (Rel-15), to be increased in subsequent releases. One of the main challenges of mMIMO is the high amount of *channel state information* (CSI) required at the BS (Marzetta 2006), which poses many new technical concerns, requiring different solutions in FDD or TDD modes. Moreover, due to the reuse of a finite number of orthogonal pilot signals in different cells, CSI acquisition is impaired by the *pilot contamination* problem in multi-cell mMIMO environments (Jose, 2009). These points led to a complete redesign of the set of reference signals used in NR for CSI acquisition and reporting.

A second aspect is that, at MMW frequencies, the transmissions will be highly directional compared to lower frequencies, i.e., they will be carried out by means of narrow TX/RX beams. A problem arising in practice is how to establish, track and possibly reconfigure such beams as the UE moves, or even when the UE device is simply rotated. In order to deal with this issue, a new set of procedures, referred to as *beam management* techniques, have been introduced in NR specifications, aimed at supporting possible fast beam reconfiguration and tracking, preferably working at the lower layers (Layers 1/2) of the protocol stack. More generally, this aspect led to the adoption in NR of a *beam-centric* approach when designing all the lower layer functions: in particular, unlike LTE, not only the user-plane channels, but also the control-plane channels can be beamformed.

Finally, at MMW frequencies, it is impractical to use conventional digital beamforming schemes for mMIMO, since their implementation requires one dedicated radio frequency (RF) chain per antenna element, which is prohibitive from both cost and power consumption perspectives. Therefore, in 5G NR specifications, support for both analog and hybrid beamforming architectures (5GRef024) is planned.

The rest of the article is organized as follows. A general introduction to multiantenna processing, which includes channel estimation issues and codebook-based precoding techniques, is provided in Section [A]. Architectural issues pertaining to multiantenna design, including digital/analog/hybrid beamforming, are discussed in Section [A]. Multiantenna solutions adopted in LTE are briefly reviewed in Section [A], while Section [A] discusses in detail the improvements carried out in 5G NR. In Section [A], a description of the new beam management functionalities of 5G NR systems operating at MMW is presented. Finally, research and standardization directions for future 5G releases as well as beyond-5G (B5G) and sixth Generation (6G) systems are contained in Section [A].

## [A] Fundamentals on multiantenna techniques

Many concepts of interest for multiantenna processing in LTE and NR can be introduced with reference to the simple MIMO narrowband model, encompassing $N_T$ TX and $N_R$ RX antennas, respectively:

$$\mathbf{y} = \mathbf{H}\,\mathbf{x} + \mathbf{n} \tag{1}$$

In (1), $\mathbf{y}$ is the $N_R \times 1$ RX vector, $\mathbf{x}$ is the $N_T \times 1$ TX vector, $\mathbf{H}$ is the $N_R \times N_T$ MIMO channel matrix, and $\mathbf{n}$ is a $N_R \times 1$ disturbance vector. Such a model is typical of a MIMO flat-fading channel, but can also represent a single subcarrier of a multicarrier transmission over a MIMO frequency-selective channel, i.e., the typical MIMO-OFDM setting in LTE and NR. Let $N_L$ be the number of layers to be transmitted, the $N_L \times 1$ vector $\mathbf{s}$ of complex symbols is mapped to the antenna vector $\mathbf{x}$ by a LP strategy, described by $\mathbf{x} = \mathbf{W}\,\mathbf{s}$, where $\mathbf{W}$ is a $N_T \times N_L$ *precoding matrix*. It should be observed that (1) can be used, with suitable interpretation of the relevant quantities, to model both a SU-MIMO system, wherein the layers in $\mathbf{s}$ are destined to a single RX, and a MU-MIMO one, where $\mathbf{s}$ gathers symbols to be delivered to different RXs (Spencer, 2004; Gesbert, 2007). Moreover, (1) can be similarly adapted to describe both DL and UL operations.

In the following, for the sake of simplicity, unless otherwise stated, we will consider the DL of a SU-MIMO system. Moreover, we will assume that the receiving UE perform linear processing by means of a $N_L \times N_R$ *decoding* or *combining matrix* $\mathbf{F}$ to recover in $\mathbf{z}$ the transmitted layers:

$$\mathbf{z} = \mathbf{F}\,\mathbf{y} = \mathbf{F}\,\mathbf{H}\,\mathbf{W}\,\mathbf{s} + \mathbf{F}\,\mathbf{n} \tag{2}$$

Design of matrices $\mathbf{F}$ and $\mathbf{W}$, either separately or jointly, is a problem that has been widely discussed in the technical literature (Oestges, 2007; Palomar, 2013) under different assumptions, by maximizing different metrics and under different constraints (e.g., per-antenna or overall power constraints at the TX side). In the simplest cases the resulting structures boil down to well-known transmitters/receivers. For instance, for a single-antenna RX ($N_R = 1$) and when the TX transmits a single layer ($N_L = 1$), equation (1) defines a MISO system with channel $\mathbf{H} = \mathbf{h}^H$. Under the assumption that vector $\mathbf{n}$ is modeled as spatially-white noise, the precoder can be simply designed by maximizing the SNR, leading to $\mathbf{w} = \mathbf{h}$, which is the *maximal ratio transmitter* (MRT). With similar reasoning, the expression of the *maximal ratio combiner* (MRC) for the decoder design can be obtained when the multiple antennas are at the RX end.

More complicated designs can be devised in the most general $N_T \times N_R$ MIMO case, with $N_L > 1$. For instance, assuming that $N_T \leq N_R$ and that the rank of $\mathbf{H}$ is equal to $N_T$ (rich scattering environment), $N_R$-fold array and diversity gains with $N_L = N_T$ spatially-multiplexed streams can be achieved by decomposing the MIMO channel into parallel independent channels (Oestges, 2007). Let $\mathbf{H} = \mathbf{UDV}^H$ be the singular value decomposition (SVD) of $\mathbf{H}$, where $\mathbf{U}$ and $\mathbf{V}$ are $N_R \times N_T$ and $N_T \times N_T$ semi-unitary matrices, and $\mathbf{D}$ is the diagonal matrix containing the nonzero singular values of $\mathbf{H}$, the precoding matrix $\mathbf{W}$ is set equal to $\mathbf{V}$, whereas $\mathbf{U}^H$ is used as decoding matrix $\mathbf{F}$. So doing, the input–output relationship (2) ends up to $N_T$ parallel independent *single-input single-output* (SISO) channels and, therefore, the mutual information of the MIMO channel turns out to be the sum of the SISO channel capacities.

The aforementioned examples show that, in order to synthesize a working TX/RX pair, the channel must be known at both TX/RX ends. Acquiring CSI at the RX (CSIR) is a standard functionality in communication systems, leveraging robust training-based channel estimation procedures (5GRef003); knowledge of the channel at the TX, so called CSIT, is more difficult to obtain. Indeed, multiantenna transmission techniques are classified as *closed-loop* or *open-loop*, depending on the availability of CSIT. Open-loop techniques, such as the celebrated space-time Alamouti scheme (Alamouti, 1998), do not require CSIT, at the price of providing only diversity gain. Closed-loop techniques exploit CSIT to obtain better performances, unleashing the full potential of spatial processing especially for beamforming, SM, and interference mitigation. Closed-loop techniques typically work better in low-mobility scenarios, where it is easier to follow the variations of the channel.

There are two types of CSI: *instantaneous* (or *short-term*) CSI and *statistical* (or *long-term*) CSI. The former amounts to know the channel impulse response (or a parametric representation of it), whereas in the latter only some statistical properties of the channel are known, such as the fading distribution, the average channel gain, the spatial correlation of the channel, and so on.

CSIT can be acquired by the BS using essentially *two* different mechanisms:

- *UE feedback*: the BS sends channel sounding signals (training symbols) to the UE, which use them to perform channel estimation, feeding back the estimation results to the BS on control channels;

- *DL/UL reciprocity*: the BS performs UL channel estimation using UE training, and adopts this knowledge as a replacement for actual CSIT in DL, exploiting the *short-term reciprocity* between DL and UL channels.

UE feedback assures very accurate and reliable CSIT if the latency of the overall process does not exceed the coherence time of the wireless channel, and offers very robust operation in cases where UL coverage is a limiting factor. In scenarios with good UL coverage, instead, DL/UL reciprocity is a viable strategy only in TDD systems, since the typical time spacing between DL/UL transmissions in such systems is smaller than the coherence time of the wireless channel, even when high terminal speeds are considered. It is worthwhile to note that reciprocity involve not only the wireless channel, but also the TX/RX chains, which typically require a calibration procedure to compensate for hardware asymmetries (Jiang, 2018). In FDD systems, short-term DL/UL channel reciprocity does not hold, due to the generally high spacing between the carrier frequencies used for DL and UL. However, also in this case some long-term channel knowledge (such as, e.g., the dominant directions-of-arrival) can be obtained by suitable averaging of UL channel estimated statistics.

The number of parameters required to describe a (possibly multiuser) MIMO channel grows with the product of the number of TX antennas, RX antennas, delay spread, and number of users. Hence, in systems resorting to UE feedback, transmitting full CSI in UL is impractical, involving an excessive use of UL control channel resources. To reduce this complexity burden, some form of *quantization* of the feedback CSI information can be adopted, leading to *quantized feedback* or *limited feedback* approaches (Love, 2008). Moreover, instead of quantizing and feeding back the raw channel parameters, a more efficient strategy is to have the RX, after acquiring the CSI, select the best precoder in a restricted set, called the *codebook*, perfectly known to both the RX and TX. Thus, only the index of the chosen precoder is fed back to the transmitter through UL control channels: if the cardinality of the codebook is kept reasonably small, this reduces to transmitting very few bits on the feedback channel. This approach, called *codebook-based precoding* (CBP), has been extensively adopted in LTE and NR as well.

A crucial point in CBP is the design of suitable codebooks. Several research studies have been dedicated to this problem and many solutions have been proposed (see Love 2008 and references therein). For the sake of implementation simplicity, the LTE standard (and later NR) adopts a DFT-based codebook (Love, 2003), whose codewords are obtained as a permutation of the columns of an (oversampled) DFT matrix. In (Love, 2003) it is proven that if the codebook contains the columns of a DFT matrix, the multiantenna system exhibits full diversity when transmitting over memoryless and independent identically-distributed (i.i.d.) MIMO Rayleigh fading channels. Later in (Hanzo, 2010) it has been proven that the DFT-based codebook approximately match the distribution of the optimal beamforming even in spatially correlated channels.

## [A] Multiantenna architectures

To implement in practical systems the multiantenna techniques described in the previous section, three prominent architectures can be considered:

- *digital beamforming*: spatial processing is performed within the digital part of the transceiver, i.e., before the digital-to analog converter (DAC) at the TX side, or after the analog-to-digital converter (ADC) at the RX side;

- *analog beamforming*: spatial processing is performed within the analog part of the TX/RX (after DAC at the TX side, before ADC at the RX side);

- *hybrid beamforming* (5GRef024): spatial processing is splitted between the digital and analog domains, which is obtained by using a number of RF chains that is strictly smaller than the number of antennas.

To better clarify the distinction, with reference to the MIMO-OFDM PHY air interface typical of LTE/NR, in a digital beamforming architecture (Figure 1) precoding is carried out digitally (in the frequency-domain) before OFDM modulation at the TX and after OFDM demodulation at the RX, whereas in an analog beamforming architecture (Figure 2) precoding is performed analogically (in the time-domain) after OFDM modulation at the transmitter and before OFDM demodulation at the receiver. Both *fully-connected* (Figure 3) and *partially-connected* (or *sub-connected*) architectures can be considered for hybrid beamforming: in the former, each RF chain is connected to all antennas, whereas in the latter each RF chain is connected only to a subset of antennas, forming a subarray.

Analog beamforming architectures typically employs only one ADC/DAC and a number of per-antenna simple analog phase-shifter and/or variable gain amplifiers. They are easier to implement especially at higher frequencies, such as MMW. However, they exhibit limited performance and flexibility, since spatial processing is applied to the RF signal with simple frequency-independent shifts. Indeed, it is not possible to perform different spatial processing across the signal bandwidth and, moreover, it is not possible to combine spatial selectivity (e.g. in MU-MIMO scenarios) with frequency-division based access, but transmissions to different users must be separated in time. However, since some spatial characteristics of the channel, such as the main directions-of-arrival or directions-of-departure, are not frequency-dependent, analog beamforming is suited to scenarios where the energy comes from a small number of dominant directions, including the line-of-sight (LOS) scenario.

Digital beamforming is much more flexible, allowing complicated frequency-dependent SM and scheduling, in order to exploit both spatial and frequency resources to operate in MU-MIMO scenarios. In particular, in a digital beamforming architecture, it is possible to apply different precoding weights for each subcarrier of a MIMO-OFDM systems (*frequency-selective precoding*), which allows one to tackle the more challenging non-LOS (NLOS) scenarios. Besides hardware constraints, the main limitation of frequency-selective precoding is the huge amount of CSI required, especially in FDD systems relying on UE feedback: for this reason, several subcarriers belonging to a subband can be grouped and precoding is performed *per-subband* rather than *per-subcarrier*.

Fully digital beamforming requires a complete RF chain (including ADC/DAC converters) for each antenna, which implies an unaffordable complexity and cost, especially for mMIMO systems and/or working at MMW (Molisch, 2017). In particular, in the latter case the large foreseen bandwidth requires high-sampling rate ADC/DAC, resulting in high power consumption and heat generation. Hybrid beamforming represents a compromise between the extreme cases of fully digital and analog beamforming (Sohrabi, 2016). Theoretical interest in hybrid beamforming is motivated by the fact that the number of RF chains can be kept small, being lower-bounded by the number of transmitted layers, while the array and diversity gain are related to the (typically much larger) number of antenna elements (Molisch, 2017). In particular, in a hybrid precoder with $N_L$ layers and $N_{RF} < N_T$ RF stages, the precoder $\mathbf{W}$ in equation (1) can be decomposed as $\mathbf{W} = \mathbf{W}_A \mathbf{W}_D$, where the $N_{RF} \times N_L$ matrix $\mathbf{W}_D$ operates in the digital domain, whereas the $N_T \times N_{RF}$ matrix $\mathbf{W}_A$ operates in the analog domain, and is mainly composed of analog phase shifters.

Since the number of BS antennas can be very large (e.g., in a mMIMO systems), the fully-connected architecture requires many analog components (phase shifters, dividers/adders) that introduce RF losses. The sub-connected architecture, instead, is expected to be more energy-efficient and easier to be implemented in MMW MIMO systems. It is worthwhile to note that particular partially-connected hybrid structures, referred to as *multi-panel* arrays, have been extensively evaluated and discussed within 3GPP during NR standardization (Huang, 2018).

A theoretical challenge in hybrid beamforming is how to design the analog and digital precoders. In principle, they should be designed jointly, for the highest performance; suboptimal approaches separate the designs for the digital and analog part. Generally, incorporating into precoding design the constraints due to the new hybrid structure (e.g., constant modulus constraint in the design of analog precoding) leads to non-convex optimization problems, which do not have a known closed-form solution and can be solved only by numerical methods. A peculiar property simplifying the design at MMW frequencies is that propagation in dense-urban NLOS environments is based only a few scattering clusters, with relatively little delay/angle spreading within each cluster (Akdeniz, 2014). In this case, the MMW channel model tend to exhibit a *sparse structure* in both angle and delay domains, which can be conveniently exploited to obtain simple precoding solutions with near-optimal performance (El Ayach, 2014).

With reference to 5G NR, implementations below 6 GHz will use (similarly to LTE) fully-digital architectures, whereas, due to implementation/cost constraints, the first 5G MMW implementations are expected to be based on analog/hybrid beamforming architectures. Since in pure analog beamforming only one TX/RX beam can be formed in one direction at any given instant, support of multiple users will require rapid *beam-sweeping* procedures. This had a profound impact on NR standardization: first, all channels and signals (including control and synchronization ones) have been designed in 5G to support beamforming (*beam-centric* design); moreover, a set of *beam management* procedures have been developed to maintain beam TX/RX pair links.

## [A] Multiantenna techniques in LTE

Multiantenna techniques were introduced in 3 GPP LTE standardization already from its first release (Rel-8). They are used to provide transmit/receive diversity, beamforming, and SM-based transmission. In 3GPP LTE, multiantenna techniques are specified as *transmission modes* (ranging from TM1 to TM10 in Rel-13 for DL), which differ in terms of the considered antenna precoding, the reference signals used for demodulation, and the procedures for acquiring CSI. The main features of multiantenna techniques for the major 3GPP LTE releases are summarized in Table 2.

## [B] DL transmission

Transmit diversity techniques are based on Alamouti-like *space-frequency block coding* (SFBC) combined with *frequency-switched transmit diversity* (FSTD), supporting transmission over 2 or 4 antenna ports. Transmit diversity is the only scheme that can be applied to all the DL channels (control/data), and is used when high reliability is required and channel-dependent scheduling is not possible (e.g., for DL control channels). On the contrary, SM schemes can be applied only to data channels. A maximum of 4 layers were supported from Rel-8, whereas later releases extended such number in DL (up to 8 layers from Rel-10) and introduced SM schemes also in UL (up to 4 layers from Rel-10). In 3GPP specifications the number of layers to be transmitted with SM is referred to as the *transmission rank*.

LTE supports in DL both linear CBP as well as *non codebook-based precoding* (NCBP) schemes, where the latter have been introduced in Rel-9**.** In CBP, pilot signals called CRS (*Cell-specific Reference Signals*) are applied *after* precoding, one for each antenna port. Hence, channel estimation performed at the UE does not include the effects of precoding, which must be explicitly signaled by the BS to the UE. Since a maximum of 4 CRS can be used in each cell, CBP allows for a maximum of 4 layers. CBP can operate both in closed-loop and open-loop modes, envisaged for low- and high-mobility scenarios, respectively. In closed-loop mode, based on channel measurements performed by means of the CRS, the UE selects a suitable transmission rank and a corresponding precoder matrix, and reports such information to the BS. The BS might follow or not the recommendation of the UE, but in the latter case it must explicitly inform the UE as to which precoder matrix is being used.

Only a limited number of precoder matrices (the *codebook*) are defined (3GPP TS 36.211, 2010) for each combination of number of layers/antenna ports. The precoding matrices of the codebook have been designed to satisfy, in addition to optimality properties, a set of conditions, such as low computational complexity, constant modulus, nested property, constrained alphabet. CBP with open-loop precoding is used when reliable feedback is not available, for example in high-mobility scenarios. In this case, only the number of layers to be transmitted is reported by the UE, while the BS employs a predefined linear processing (known to the UE) composed by a combination of precoding and *cyclic delay diversity* (CDD), providing not only SM but also increased robustness thanks to diversity.

NCBP was introduced in Rel-9 (single-layer, TM7), and later extended in further releases (TM8-TM10) to allow for transmission of a maximum of 8 layers. Compared to CBP, the main difference is the insertion of UE-specific pilot signals called *Demodulation Reference Signals* (DM-RS), which are applied *before* precoding.  That is, reference signals are jointly precoded with the data, which allows coherent demodulation of the layers at the UE without explicit knowledge or indication of the actual precoding scheme applied at the BS (so called "transparent precoding"). In this case, the BS can select an arbitrary precoder and the UE only has to know the number of transmitted layers, since the precoding matrix is jointly estimated as part of the channel.

The choice of a suitable precoding matrix from the BS in NCBP can be done in two ways. A first solution, of interest for TDD operation, is based on UL/DL reciprocity: the BS acquires CSI on the basis of *Sounding Reference Signals* (SRS) transmitted by the UE and employs the estimated CSI to optimize its precoding matrix. A second solution, useful for FDD mode, is based again on CSI feedback from the UE; it should be noted that the feedback procedure is very similar to that adopted in closed-loop CBP, but the precoding codebook is only used for UL CSI reporting and not for actual DL transmission.

**[B] UL transmission**

UL multiantenna schemes were introduced only in Rel-10 (3GPP TS 36.211, 2010). Focusing on data transmission, SM precoding schemes (up to 4 layers) introduce DM-RS signals for channel estimation before precoding, similarly to DL NCBP schemes. However, in UL the precoder matrix to be used by the UE is chosen by the BS in a given codebook, so as to limit DL signaling burden. Since UL precoding occurs after DFT transform precoding in the UE, the codebook matrices are chosen so as to have at most one layer mapped to each antenna port, in order to preserve the good cubic-metric properties of the transmitted signal (so called "cubic metric preserving codebook").

## [A] NR multiantenna techniques

Most Rel-15 multiantenna techniques for NR are smooth evolution of those already presented for LTE (Rel-14). Nevertheless, support for a massive number of antennas (e.g. up to 64) as well as extensive implementation of mMIMO/FD-MIMO schemes are envisioned (5GRef023). NR do not specify precoding schemes in DL, since the precoding is NCBP (*transparent* to the UE), with a maximum of 8 transmitted layers in Rel-15. In UL, both CBP and NCBP schemes are employed, with up to 4 layers in Rel-15. These figures are not much different from those of later LTE releases (see Table 2). However, the improved CSI estimation and reporting framework of NR assures a significant gain over LTE in DL spectral efficiency.

A groundbreaking difference between LTE and NR is the *beam-centric* design adopted by the latter, which implies a redefinition of the structure and functions of the main reference signals. In particular, one of the main change in NR design has been to remove the "always-on" CRS transmitted in DL, which performs in LTE many important functions. These functions have been distributed in NR among other reference signals, which are tailored for specific roles and can be flexibly adapted for different deployment scenarios and spectrum plannings.

To support highly directional transmissions at MMW frequencies, a set of improved and flexible functionalities have been introduced in NR to support beam acquisition, tracking, sweeping, recovery, referred to as *beam management* procedures. Moreover, since at MMW frequencies phase noise is expected to adversely affect the performance (especially for larger constellation, such as 64 or 256 QAM), a specific signal called *Phase Tracking Reference Signal* (PT-RS) has been introduced to estimate and compensate for phase noise impairments (5GRef008).

Both DL and UL data transmissions are based on DM-RS for coherent demodulation. In particular, the number of layers that can be transmitted simultaneously depends on the maximum number of orthogonal DM-RS signals defined by the standard. For SU-MIMO a maximum of 8 layers are supported in DL, while a maximum of 4 layers are supported in UL. For MU-MIMO a maximum of 12 layers can be supported for both DL and UL, with up to 2 layers per scheduled UE. When the UE employs DFT-s-OFDM, only single-layer transmission is allowed in UL to preserve the cubic metric. Finally, unlike LTE, TX diversity schemes is currently not explicitly supported in NR, and can be possibly employed in a specification-transparent manner, i.e., using precoder cycling in frequency.

## [B] DL transmission

In NR, similarly to LTE, a DL *physical channel* corresponds to a set of resource elements carrying information originating from higher layers. The following channels are defined (3GPP TS 38.211, 2019):

- *Physical Downlink Shared Channel* (PDSCH);

- *Physical Broadcast Channel* (PBCH);

- *Physical Downlink Control* Channel (PDCCH).

The first channel is devoted to user data transmission, whereas the others perform control functions. Moreover, a set of downlink *physical signals* are defined, which are used to support important operations at the UE, such as the DM-RS, the PT-RS, the *Channel-State Information Reference Signals* (CSI-RS), the *Primary/Secondary Synchronization Signals* (PSS/SSS). Such signals are generated and used in the physical layer only, and do not carry information from/to higher layers. In the following we will discuss in particular the important role played by DM-RS and CSI-RS in DL multiantenna processing.

Unlike last releases of LTE, which provided 10 DL transmission modes, in NR Rel-15 there is only one DL transmission mode; however, this NR "single mode" is very flexible, supporting a vast number of new functionalities, especially those related to beam management.

Focusing on PDSCH (data channel), the coded bits to be transmitted are scrambled with the aid of a length-31 Gold sequence and modulated by one of the allowable modulation schemes (BSPK/QPSK/16QAM/256QAM). The resulting complex-valued symbols are mapped to layers (up to 8) as described in (3GPP TS 38.211, 2019), then the resulting vectors are mapped to antenna ports, and finally to virtual/physical resource blocks. PDCCH and PBCH channels are built similarly (they use QPSK modulation) and are described in detail in (3GPP TS 38.211, 2019).

Different from LTE, NR employs only NCBP schemes for DL transmission; precoding codebooks are still defined in the standard, but are used only to simplify CSI reporting. Different DM-RS signals, one for each layer, are jointly precoded with the data (Figure 4); hence, the UE is able to estimate the *composite* channel obtained by concatenating the precoding matrix chosen by the BS with the actual MIMO channel (*transparent* precoding). The choice of the actual precoding scheme in DL is not specified by the standard and is left to the manufacturer, which obviously makes DL multiantenna transmission very flexible.

The precoding matrix to be used by the BS might depend either on CSI reporting made by the UE (typical in FDD systems), or by CSI acquired by the BS itself in UL on the basis of SRS signals transmitted by the UE (typical in TDD systems). Therefore, the main aspects covered by 3GPP specifications (3GPP TS 38.214, 2019) are the measurement and reporting setting made by the UE to help BS design a particular precoder for DL data transmission.

A CSI report comprises the following items:

- *Rank Indicator* (RI), indicating a suitable transmission rank (i.e., the number of layers for SM) for DL transmission;

- *Precoder Matrix Indicator* (PMI), indicating a suitable precoder matrix;

- *Channel Quality Indicator* (CQI), indicating a suitable channel coding and modulation scheme.

Again, it should be noted that the concept of precoder codebook is only used for reporting, but do not impose any restriction on the precoder actually used by the BS for DL transmission. As a matter of fact, the BS is free to choose any precoder in order to optimize system-wide metrics, which is a common choice in MU-MIMO. Indeed, while in SU-MIMO scenarios the BS can use the precoder indicated by PMI reported by the UE, in MU-MIMO the BS generally has to optimize simultaneous MIMO transmission to multiple UE in the same time/frequency resource. In this case, the BS should perform joint optimization of the precoding matrices for all UEs scheduled for transmission, aimed not only at improving transmission to a given UE, but also at reducing interference among DL transmissions to different UEs.

CSI estimation and reporting in DL is based on CSI-RS: the UE performs channel measurement with CSI-RS and reports the selected precoding matrices to the BS for reference. How to use the reported precoding matrices for link adaptation and scheduling is left to BS implementation.

MU-MIMO operation (5GRef021) typically requires more detailed CSI compared to transmission to a single device (SU-MIMO). For this reason, two different CSI types are supported in NR (3GPP TS 38.214, 2019), which differ in the structure and size of the precoder codebooks used for reporting by the UE:

- *Type-I CSI* (standard resolution), optimized for SU-MIMO transmission with a potentially large (up to 8) number of layers;

- *Type-II CSI* (high resolution), optimized for MU-MIMO transmission, with up to 2 layers per scheduled UE and an overall maximum number of 12 layers.

Type-I CSI is similar to Class A CSI proposed in LTE Rel-13 and 14: it is a relatively simple codebook, exhibiting a small UL overhead. Instead Type-II CSI is a new feature of NR and provides finer channel information at the price of larger UL overhead: indeed, the limitation of a maximum of 2 layers/UE is mostly due to the reduction of such overhead. For this reason, some studies (Ahmed, 2019) targeted at NR Rel-16 apply some form of compression to reduce feedback overhead, with the aim of supporting a larger number of layers/UE.

The codebooks for Type-I CSI are relatively simple, being aimed at focusing the transmitted energy toward the UE by using a single beam; interference between the possibly large number of SM layers must be managed by RX processing at the UE. A *dual-stage* codebook is employed, wherein the precoding matrix **W** is expressed as the product of two matrices $\mathbf{W}_1$ and $\mathbf{W}_2$, with $\mathbf{W}_1$ capturing long-term and frequency-independent (wideband) channel characteristics, and $\mathbf{W}_2$ taking into account short-term and potentially frequency-dependent (subband) channel characteristics.

Two subtypes of Type-I CSI are *single panel* and *multi-panel CSI*: the former is designed by assuming a single antenna panel with $N_1$ x $N_2$ cross-polarized antenna elements, whereas the latter assume an antenna configuration with $N_g$ = 2 or $N_g$ = 4 two-dimensional panels, each with $N_1$ x $N_2$ cross-polarized antenna elements (Figure 5). A maximum of 32 cross-polarized antenna ports is supported by the standard, with several allowed configurations of ($N_g$, $N_1$, $N_2$). The operation principles of Type-I single-panel and multi-panel CSI are similar, except that the former supports transmission of up to 8 layers to the same UE, while the latter supports a maximum of 4 layers. Moreover, in multi-panel CSI the matrix $\mathbf{W}_1$ defines one beam per polarization and panel, whereas $\mathbf{W}_2$ provides per-subband co-phasing between polarizations as well as between panels.

The codebooks for Type-II CSI are more complicated, allowing the UE to report CSI with much higher granularity, in order to enable interference management typical of MU-MIMO. While Type-I CSI selects and reports a single beam, Type-II CSI is oriented to multiple-beam operation, allowing to report up to 4 orthogonal beams. For each selected beam/polarization, the reported PMI includes both amplitudes (wideband and per-subband) and phases (per-subband). This allows one to build a more detailed model or the channel, capturing the main rays and their respective amplitudes and phases.

At the BS, the PMI delivered from multiple devices can be used to suitably configure a MU-MIMO transmission scheme, identifying a set of UEs that can be served simultaneously on the same time/frequency resources (with a maximum of 2 layers per device) and the relative precoding matrices.

In TDD systems, DL precoding can be based on CSI acquired by the BS assuming DL/UL reciprocity. Therefore, in this case the BS perform channel estimation by using the pilot signals (5GRef003) transmitted by the UEs: similarly to LTE, in NR this procedure exploits SRS transmissions. In order to preserve the cubic metric when DFT-s-OFDM is applied at the UE, SRS signal design is based on *Zadoff-Chu* sequences (Chu, 1972), which assure good orthogonality properties in the frequency domain, as well as constant envelope in the time domain.

In (Vook, 2018) a system-level performance comparison between LTE and NR operating at 2 GHz has been carried out by numerical simulations for the same number of antenna ports (16 or 32). It is shown that the new NR Type II-CSI codebook assure significant performance gains (in terms of spectral efficiency) over the best LTE codebook, especially in MU-MIMO operation mode, with gains ranging from 19% to 35% in mean spectral-efficiency, and from 12% to 32% in cell-edge performance. Similar results are obtained in (Mondai, 2019)**,** which show in addition that the gains due to higher CSI accuracy (Type II) are significantly reduced for moderate UE speeds, and with a reduced number of antennas (8 compared to 32).

### [B] UL transmission

An UL *physical channel* is the set of resource elements carrying information originating from higher layers. The following channels are defined (3GPP TS 38.211, 2019):

- *Physical Uplink Shared Channel* (PUSCH);

- *Physical Uplink Control Channel* (PUCCH);

- *Physical Random Access Channel* (PRACH).

Of these channels, the first transports data, the second is used for control-plane functions, and the third one for initial access. Focusing on PUSCH transmission, NR supports UL SM schemes with up to 4 layers: if DFT-s-OFDM is employed, only single-layer transmission is allowed. In addition to the CBP scheme, which is an extension of the corresponding LTE scheme, NR introduces also a more flexible NCBP scheme in the UL. Similar to the DL, it is assumed in every case that any precoding is applied also to the DM-RS signals transmitted with the data, which makes UL precoding transparent to the network for both CBP/NCBP operations.

For CBP schemes, different codebooks for DFT-s-OFDM and CP-OFDM waveforms are designed (3GPP TS 38.211, 2019). In DFT-s-OFDM transmission mode, only rank 1 (single-layer) transmission over 2 or 4 antenna ports is supported, and the employed codebook is an extension of LTE Rel-10 design. For CP-OFDM, a DFT-based codebook design is adopted, similar to LTE, allowing transmission of a maximum of 4 layers over 4 antenna ports.

An important aspect that might impose constraints on UL precoding is the assumption of phase coherence between different UE antennas. Indeed, NR specifications allow for different coherence capabilities of the UE, referred to as *full coherence*, *partial coherence,* and *no coherence*. Accordingly, only a subset of the codebook can be chosen to match the UE antenna coherence capability; in particular, non-coherent UEs can only perform antenna-selection precoding, whereas partially-coherent UEs can use linear combination within pairs of antennas (with selection between the pairs), and, finally, fully coherent UEs can access the whole codebook set, with possible linear combination over all antenna ports.

In CBP, based on channel measurements performed by the BS on SRS transmissions made by the UE, the BS selects a transmission rank and a corresponding precoding matrix, considering also the device capabilities in terms of antenna-port coherence. Differently from DL, where precoder feedback is only used for reporting, in UL the UE is assumed to use the precoder indicated by the BS.

A fundamental difference between NR and LTE is that a device can transmit multiple SRS over separate and relatively wide beams (Figure 6).  For example, these beams may correspond to different device antenna panels pointing to different directions. In this case, BS feedback includes also a *SRS Resource Indicator* (SRI) identifying the SRS and hence the beam to be used for transmission, while the RI and PMI define the number of layers and the precoder to be used within the selected beam.

CBP is typically used when UL/DL reciprocity does not hold, that is, when the UE is not able to autonomously infer optimal precoding options, as in FDD mode or for TDD with non-calibrated hardware. NCBP is used instead when the UE can assume reciprocity (in TDD mode) and acquire CSI by performing DL measurements on CSI-RS transmitted by the BS.

In NCBP mode, based on DL measurements carried out on CSI-RS configured by the BS, the UE selects a suitable UL precoder, without being restricted to a particular codebook. Each column of the precoder matrix **W** defines a digital beam to the BS for the corresponding layer. Since the precoder choice made by the UE might not be optimal from the BS point of view, the BS can modify the precoder chosen by the UE, by removing some beams or equivalently some columns from the selected precoding matrix.

To enable this, the device (Figure 7) first applies the selected precoder to a set of configured SRS, one for each beam defined by the precoder. Based on SRS measurements, the BS indicates to the UE a subset of the configured SRSs by using the SRI. The UE then carries out the transmission using a modified precoder, where only the columns (beams) indicated by the SRI are included.

## [A] Beam management

NR working at FR2 frequencies (MMW) will make extensive use of analog or hybrid beamforming to overcome the path loss penalty and achieve improved spatial selectivity by means of highly directional transmissions. At MMW frequencies, the quality of signal beams is affected even by small movement of the device or body obstructions, which might lead to rapid drops in signal strength. Therefore, NR must provide efficient and fast mechanisms to establish and adaptively manage highly directional links, which are collectively known as *beam management* (3GPP TR 38.912, 2018; Giordani, 2019) procedures.

The main task of beam management is to acquire and maintain a reliable beam pair, that is, a TX beam and a corresponding RX beam that jointly provide good radio connectivity. The following beam management procedures are defined in (3GPP TR 38.912, 2018), which are used to perform *initial access* for idle users and *adaptation/recovery* for connected users:

- *beam sweeping*: the BS or UE covers a large spatial area with a set of beams transmitted and/or received during a time interval in a predetermined way;

- *beam measurement*: the BS or UE measures the quality of the received beamformed signals by using the received power or more sophisticated metrics, such as the SNR or the signal-to-interference-plus-noise ratio (SINR);

- *beam determination*: the BS or UE selects one or multiple beams to ensure good radio link quality;

- *beam reporting*: the UE reports information regarding the beamformed signals on the basis of beam measurements.

Beam management employs fast procedures, involving only Layer 1/2 signaling, which work by sending reference signals (SS blocks or CSI-RS in DL, SRS in UL) in a number of candidate beams, and estimating the quality of the reference signal at the RX for each beam. In many cases, a suitable TX/RX beam for the DL is also a suitable beam pair for the UL and vice versa. This is a form of spatial reciprocity, referred to in NR as *beam correspondence*. Since beam management does not deal with small-scale channel variations, beam correspondence can be applied also to FDD mode.

**[B] Initial access**

At low frequencies, control channels for initial access can be transmitted with a wide beam (e.g., covering an entire sector/cell). At MMW frequencies, however, also control channels for initial access need to be beamformed to achieve the necessary coverage. Since the direction to the UE is not known in advance, *beam sweeping* over the entire cell sector must be adopted. It should be mentioned that in NSA deployments initial access at MMW could be simplified by dual-connectivity and resorting to the LTE control-plane (Giordani, 2019).

The reference signal used for initial access is the SS block transmitted by the BS. Up to 64 beams can be transmitted within an SS burst of 5 ms, which is repeated every 20 ms. The UE (Figure 8) perform measurements on the different SS blocks to determine the best beam, and use the PRACH channel associated to the beam to access the network and indicate to the BS which is the best beam for subsequent DL data/control transmission or for possible beam refinement procedures. The SS block beams are relatively wide, to provide robustness against blockage and mobility and reduce signaling overhead. These wide beams could be sufficient to perform low data-rate transmission to the UE. Otherwise, more directional beams can be acquired by *beam refinement* procedures based on UE-specific CSI-RS signals.

### [B] DL beam refinement/adaptation

After initial access, the BS might perform beam refinement (Figure 8) by transmitting a set of beamformed CSI-RS to the UE in a narrow angular sector centered on the wide beam acquired from SS block during initial access. The UE measures again the quality of the received beams and reports the results to the BS. The UE might report a single beam or a group of BS beams that can be received simultaneously by the UE, e.g., by using different antenna panels. Based on UE reports, the BS determines the TX beam (or a group of beams) to be used for subsequent DL transmission and indicates such TX beam(s) to the UE so that proper RX beam(s) can be applied.

The beam characteristics should be regularly adjusted due to movements and/or rotations of the mobile device, or even due to modifications of the propagation scenario. This is usually done in two steps: TX beam adjustment, which refines the TX beam while keeping the RX one fixed, and RX beam adjustment, which refines the RX beam while keeping the TX one fixed.

In the first step, the UE reports measures of different DL beams to the BS, which may decide to adjust the current TX beam(s). In the second one, the UE performs measurements of different RX beams obtained by sweeping; based on such measurements, beam adjustment can be carried out directly by the UE, without any reporting or intervention by the BS.

### [B] DL beam recovery

Due to mobility and blockage, the current beam pair between the BS and UE may be blocked, resulting in a *beam failure* event. Beam failure could lead to *radio link failure* (RLF) already defined in LTE, which is managed by a costly higher-layer reconnection procedure. Since beam failures might occur rather frequently, NR supports a faster *beam recovery* procedure using lower-layer signaling.

Beam-failure detection is based on measurements of the quality of some reference signal (SS block or CSI-RS). To find a new beam, the UE monitors SS blocks or CSI-RS over a number of candidate beams. When a new beam has been found, the UE issues a *beam-recovery request* to the serving BS on the PRACH. If the response is not received within a certain timeout, RLF is declared and higher-layer reconnection procedures are triggered.

### [B] UL beam management

UL beam management can be done in two different ways. When beam correspondence holds and a DL beam has been established, explicit UL beam management is not needed.  Otherwise, if explicit UL beam management is needed, it can be performed in essentially the same way as in DL, using the SRS signal instead of the CSI-RS or SS block.

### [A] 5G evolution, emerging technologies and conclusions

As of December 2019, 5G Rel-16 (informally known as "5G Phase 2") is in the final stage, which is forecasted to be frozen in March 2020 and completed in June 2020.  Useful details about the main features will be contained in (3GPP TR 21.916, 2018) which, at the time of writing, is not publicly available. We will thus rely on some recent papers (Dahlman, 2019; Ghosh, 2019; Kim, 2019) for a brief discussion about the expected improvements, with focus on those involving or related to multiantenna processing.

Rel-16 is a major release, as it will meet the challenging ITU IMT-2020 requirements (ITU-R M.2083, 2015). Innovations will be both at the radio level, i.e., operation for unlicensed spectrum (NR-U), integrated access/backhaul (IAB), integration of satellite access, as well as the support of new applications, such as industrial IoT, V2X services, and mission-critical applications. Performance enhancements of eMBB MIMO and multiantenna techniques will be featured, including enhanced MU-MIMO support based on enhanced CSI feedback, enhanced multi-panel transmission, and enhanced multi-beam operation (Dahlman 2019).

In particular, Rel-16 will address the high overhead problem of Type II CSI feedback, by introducing new compression techniques, but also extending the applicability of Type II CSI to a maximum of 4 layers per-user in MU-MIMO scenarios. Moreover, support for transmission and reception at multiple points will be introduced, enabling coordinated multipoint (CoMP) operation, based on *non-coherent joint transmission* (NC-JT). This will be especially relevant for ensuring the desired reliability for URLLC services. Finally, beam management operation will be extended to handle more than 64 beams with a reasonable signaling overhead.

Work for Rel-17 is still in an early phase, targeting availability of the new specification in mid-2021. In June 2019, the 3GPP community identified the main topics of interest, and a final set of Rel-17 features will be selected in December 2019. Further MIMO enhancements are scheduled, motivated by current commercial deployments, as well as support for cases with high-speed mobility, and better support for FDD operation. An interesting development is operation over 52.6 GHz, where more efficient modulation schemes than OFDM, such as single carrier modulation, might be considered,  to cope with the limitations of power-amplifier technologies at such high frequencies. With reference to multiantenna aspects, it is expected that in Rel-17 further enhancements to beam management and handling of path diversity (to compensate for the blockage phenomenon) will be studied. Moreover, with reference to MIMO technologies, techniques for reducing the CSI overhead and/or to partially exploit reciprocity even in FDD operations will be investigated. Moreover, it is likely that Rel-17 will address UL-MIMO operation with additional support for NCB operation and enhancements to CBP operations (Ghosh, 2019).

The feeling arising from the study of the standardization process of NR is that there is no single enabling technology that can support all 5G application requirements. New user requirements, new applications and use cases, and new networking trends will bring more challenging communication paradigms, especially at the PHY layer. In this respect, emergent technologies for B5G and 6G systems will attempt to exploit the implicit randomness of the propagation environment in order to either simplify the transceiver architecture and/or to increase the quality of service (QoS). Notable examples are briefly discussed in the forthcoming subsections.

**[B] Spatial modulation**

Conventional MIMO schemes rely either on SM to boost the data rate or diversity to improve the bit-error-rate performance. The TX antennas of a MIMO system can be used to convey additional information bits by means of a novel digital modulation technique called *spatial modulation* (Basar, 2016), which is based on the simple idea of altering the on/off status of the antennas at the transmitter. Specifically, in a multiantenna system with $N_T$ TX antennas that employ conventional M-ary signal constellations, such as PSK or QAM, a simple form of spatial modulation consists of transmitting $\log_2(M)+\log_2(N_T)$ bits in each signaling interval, where $\log_2(M)$ bits are used to modulate the phase and/or the amplitude of a carrier signal, while the remaining $\log_2(N_T)$ bits are reserved to select the index of the active TX antenna that transmits the modulated signal. At the RX side, the maximum likelihood detector can jointly search for all possible TX antennas and M-ary constellation symbols to decide on both the transmitted symbol and the index of the activated transmit antenna. Compared to a classical MIMO system, spatial modulation allows to conveying information in a more energy-efficient way without increasing the hardware complexity.

During 3GPP RAN1#87 meeting in November 2016 and 3GPP TSG RAN WG1 NR Ad-Hoc Meeting in January 2017, it has been proposed to further evaluate the adoption of spatial modulation for 5G NR. As an evolution of the basic concept of index modulation for transmit antennas, *spatial scattering modulation* (Ding, 2017) and *beam index modulation* (Ding, 2018) exploit the indices of the scattering clusters available in the environment to convey information. Furthermore, media-based modulation utilizes reconfigurable antennas (Basar, 2019) by encoding the information bits onto multiple distinguishable radiation patterns.

### [B] Nonlinear precoding

LP schemes, which are widely studied in the literature and integrated into current 3GPP standards, may not offer good performance in high-UE density scenarios, which can exhibit strongly correlated channels. For this reason, a great bulk of research has been recently devoted to *nonlinear precoding* (NLP) schemes for massive MU-MIMO systems, employing nonlinear processing to mitigate interference. NLP schemes are indeed more robust to channel correlation among users, and are potentially capable to enhance the MU-MIMO systems performance in 5G NR.

Research and standardization activity on NLP within 3GPP are summarized in (Hasegawa, 2018). Viable NLP strategies are *Dirty-Paper Coding* (DPC) (Costa, 1983), which performs a pre-subtraction of the non-causally known interference, or *Tomlinson-Harashima Precoding* (THP) (Tomlinson, 1971; Harashima, 1972), which represents a reduced-complexity alternative to DPC. NLP integration into practical systems requires significant efforts, especially in terms of sensitivity to CSI errors and high computational complexity. Differently from LP techniques, usually relying on subspace approaches, THP resorts to successive interference cancellation to suppress multiuser interference. This mechanism renders THP highly susceptible to CSI inaccuracies due to estimation errors, limited feedback or feedback delay. In the 3GPP community, also hybrid solutions that combine NLP and LP have been considered; in particular, NLP and LP can be used jointly or switched dynamically.

In the scientific literature, some recently proposed solutions rely on two-stage structures (Begashow, 2019; Adhikary, 2013), where a *joint spatially-division and multiplexing* (JSDM) approach is pursued to enhance the MU-MIMO performance and, at the same time, simplify system operations also for FDD systems. In particular, since the channels between BS and UEs are usually correlated depending on scattering geometry, the UEs can be partitioned into groups with approximately the same channel covariance eigenspace. Thus, the resulting DL beamformer can be split into two stages (Adhikary, 2013): a first stage, called *pre-beamforming* (PBMF), which depends only on the second-order statistics of the channel, and a second stage, called *precoding stage*, which relies on the instantaneous realization of the effective channel, including also the pre-beamforming processing. Such a precoding stage is chosen as to minimise the intergroup interference. The second stage needs, of course, estimation and feedback of the instantaneous effective channel, but this operation entails a reduced complexity thanks to user partitioning.

Interestingly, it is shown that, in the case of uniform linear arrays, the PBMF matrix can be obtained by selecting blocks of columns of a unitary DFT matrix. DFT-based PBMF achieves very good performance and effective channel dimensional reduction and requires only a rough knowledge of the DOA distribution for each user group, without requiring an accurate estimation of the channel covariance matrix. In (Zarei, 2016), a two-stage precoding architecture is proposed, where users are partitioned into groups based on similarity of their channel correlation matrices, and block diagonalization precoding is used to suppress the intergroup interference, while THP is employed to mitigate the intragroup interference.

**[B] Reconfigurable intelligent surfaces**

As previously pointed out, NR and beyond is expected to migrate to higher frequencies, e.g., the MMW (FR2) and even sub-MMW (above 100 GHz) bands (Rappaport, 2019). At these frequencies, signals are highly susceptible to blockages from large-size structures on the radio path, e.g., buildings, and they are severely attenuated by the presence of small-size objects, e.g., human bodies and foliage. Such problems become more pronounced in dense urban environments due to the highly dynamic nature of the radio environment. A possible approach to circumvent the unreliability of high-frequency channels is to artificially create additional routes between the source and the destination. Along this line, the solution adopted in LTE-Advanced (LTE Rel-10) is based on the deployment of relay stations that capitalize on the concept of *distributed cooperative diversity* (Laneman, 2004).

*Relay nodes* are low-power BSs that provide enhanced coverage and capacity at cell edges and hotspot areas, and can also be used to connect to remote areas without fiber connections. However, the use of relay BSs usually reduces the network energy efficiency and increases network complexity, while requiring a larger capital expenditure for deployment. Up to now, the new generations of wireless networks have been developed under the well-accepted idea that higher data rates come at the cost of increased power consumption and radio-wave emission. In contrast, a clear trend that seems to emerge for B5G and 6G networks is to improve system performance without increasing power consumption, but "recycling" signals generated by legacy transmitters. According to this vision, when the LOS path does not exist or is of insufficient quality, a viable alternative to relaying is using *reconfigurable intelligent surfaces* (RIS).

RIS are artificial 2D (planar) or 3D (spherical) surfaces that are capable of altering the propagation of the radio waves impinging upon them. RIS can be based on *meta-surfaces* consisting of subwavelength structures, which are also known as *meta-atoms*, whose electromagnetic properties can be controlled with integrated electronics. The subwavelength structures, which are typically far thinner than the working wavelength, can abruptly change the phase, amplitude, and polarization state of the incident RF signal. As a consequence of such a discontinuity, the refracted and reflected beams from RIS follow the *generalized Snell's laws* (Yu, 2011), according to which the angle of refraction/reflection can be different from the angle of incidence (not predicted by the ordinary Snell's law, which is based on the assumption that no abrupt changes occur over the scale of the wavelength). This phenomenon provides great flexibility in beamforming and shaping radio waves in a real-time controlled manner, by providing unprecedented energy-focusing, data-transmission, and terminal-positioning capabilities. Specifically, the main distinguishable features of RIS compared to related technologies currently employed in wireless networks, such as relaying and MIMO beamforming, are the following ones:

- nearly passive without requiring dedicated energy sources;

- not affected by receiver noise (ADC/DAC converters and power amplifiers are not required);

- inherently full-duplex;

- easy to be deployed (e.g., on the facades of buildings, ceilings of factories and indoor spaces, human clothing, etc.);

- software-controlled (possibly based on artificial intelligence).

By adaptively changing the states and phases of the meta-atoms, RIS allow to implement *reflect beamforming* or *passive beamforming*, by achieving coherent superposition of the reflected signals at a desired RX, improving thus the received SNR. However, RIS not only can be used to improve the SNR at the RX, but also to deterministically control the propagation environment. This is a radical different viewpoint with respect to current communications models according to which the channel is seen as an uncontrollable and hostile system, whose adverse effects have to be counteracted by judiciously designing TX and RX such that to adapt themselves to the radio environment.

In contrast, RIS play an active role in transferring and processing information, thus making the channel dynamically reconfigurable by system designers. From a model viewpoint, this means that the overall system, i.e., TX, channel, and RX can be jointly optimized to further improve the network performance, by allowing the transmitter control the state of RIS via a finite-rate control link.

In summary, RIS and, more in general, *smart radio environments,* allow to translate network softwarization from the logical domain into the physical domain, by seeing the radio environment as a software entity that can be remotely programmed, reconfigured, and optimized. Along this vision, many PHY solutions in 5G NR, including beamforming and precoding, are envisioned to be redesigned in future releases.

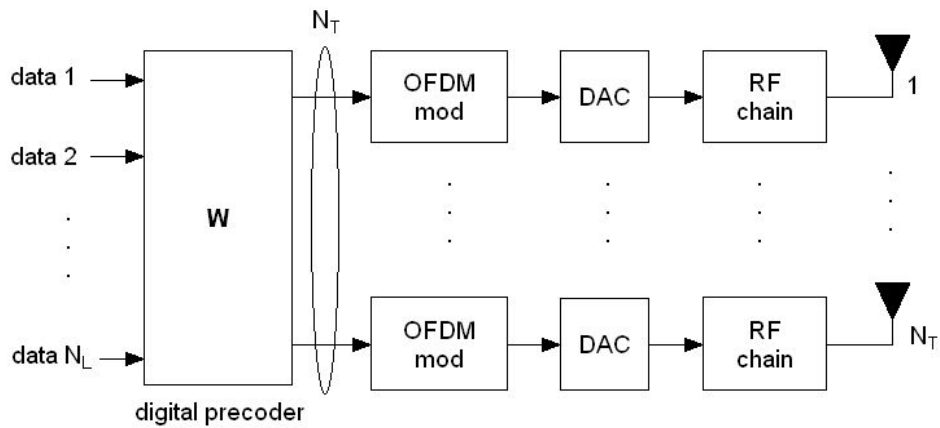

**Figure 1.** Digital beamforming architecture for a multiantenna OFDM transmitter.

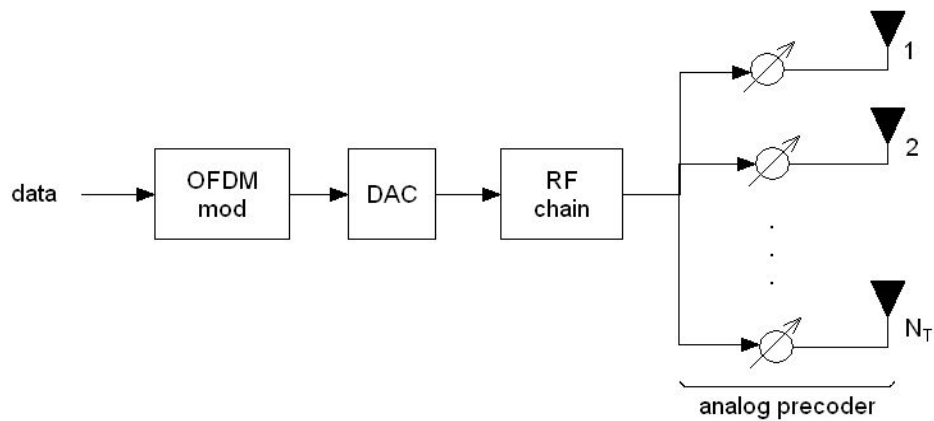

**Figure 2.** Analog beamforming architecture for a multiantenna OFDM transmitter.

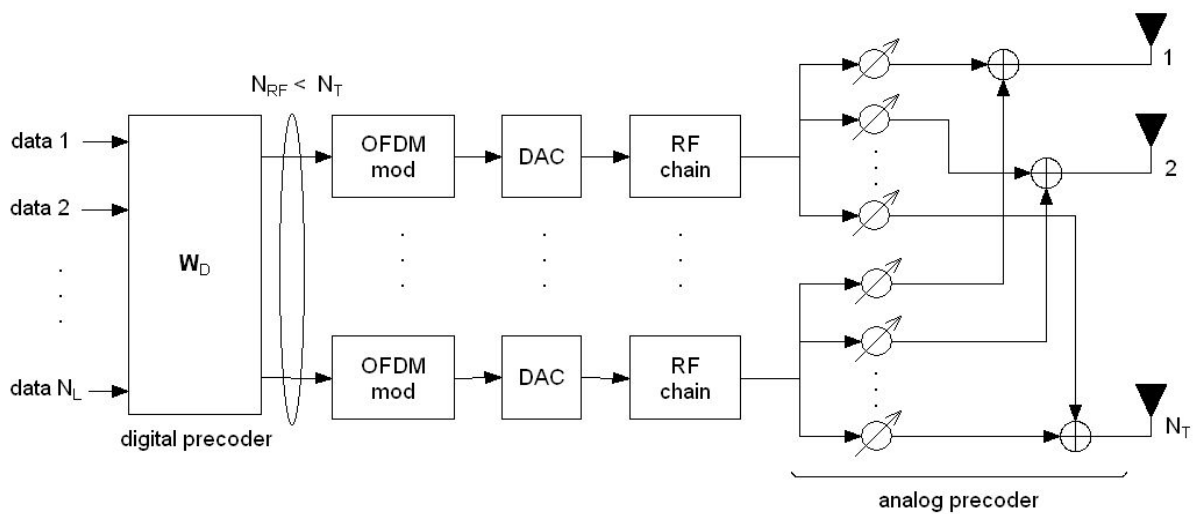

**Figure 3**. Hybrid beamforming (fully-connected) architecture for a multiantenna OFDM transmitter.

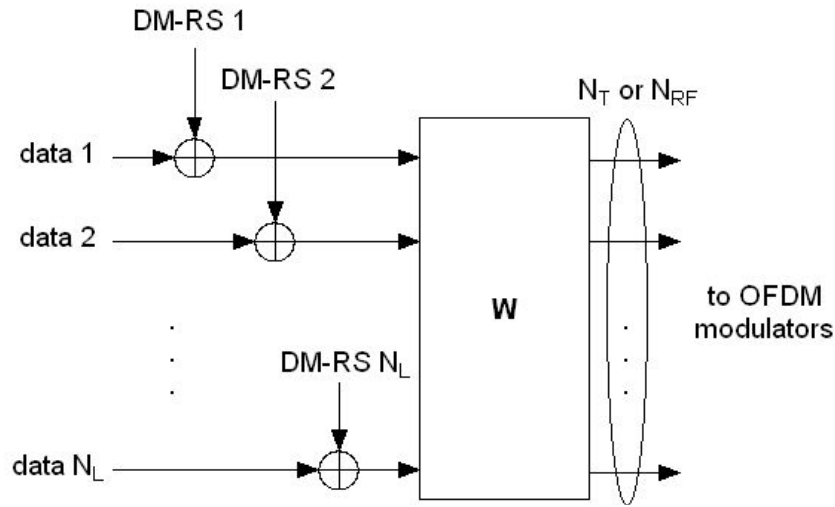

**Figure 4.** DL "transparent" precoding scheme for NR, where the DM-RS signals undergo the same precoding as data. The number of outputs is $N_T$ or $N_{RF}$ depending on whether a digital or a hybrid beamforming architecture is employed.

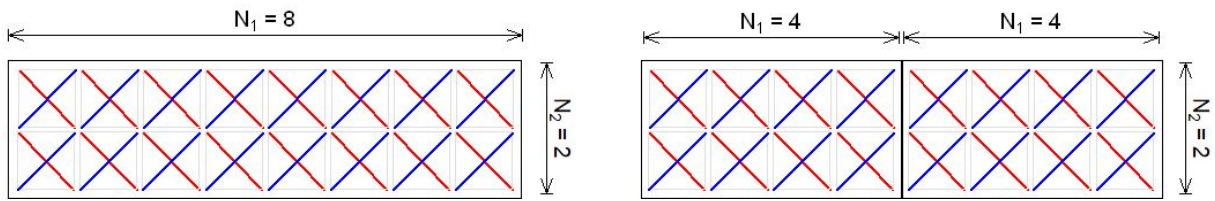

**Figure 5.** Single-panel (left) and multi-panel (right) 32-port cross-polarized antenna configurations. Each antenna element encompasses two orthogonal polarizations (indicated with red/blue colors).

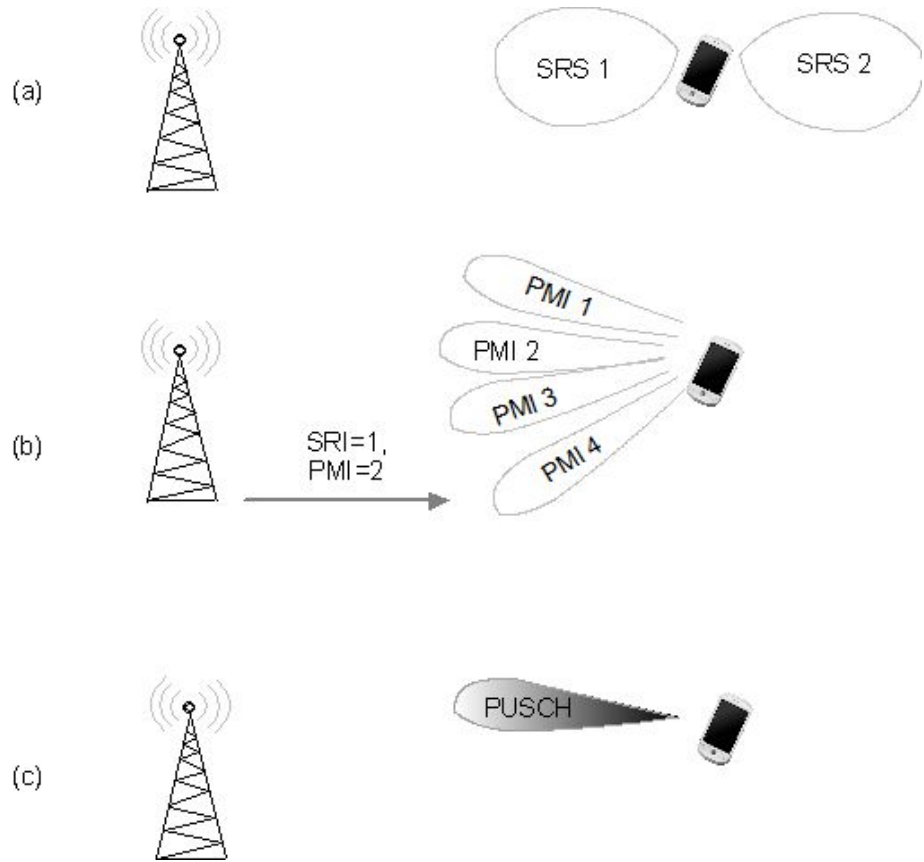

**Figure 6.** Codebook-based UL transmission in NR. (a) The UE transmits two wide SRS beams for UL channel estimation; (b) the BS determines the best SRS and a suitable UL precoding schemes; (c) the UE uses the indicated beam for PUSCH transmission.

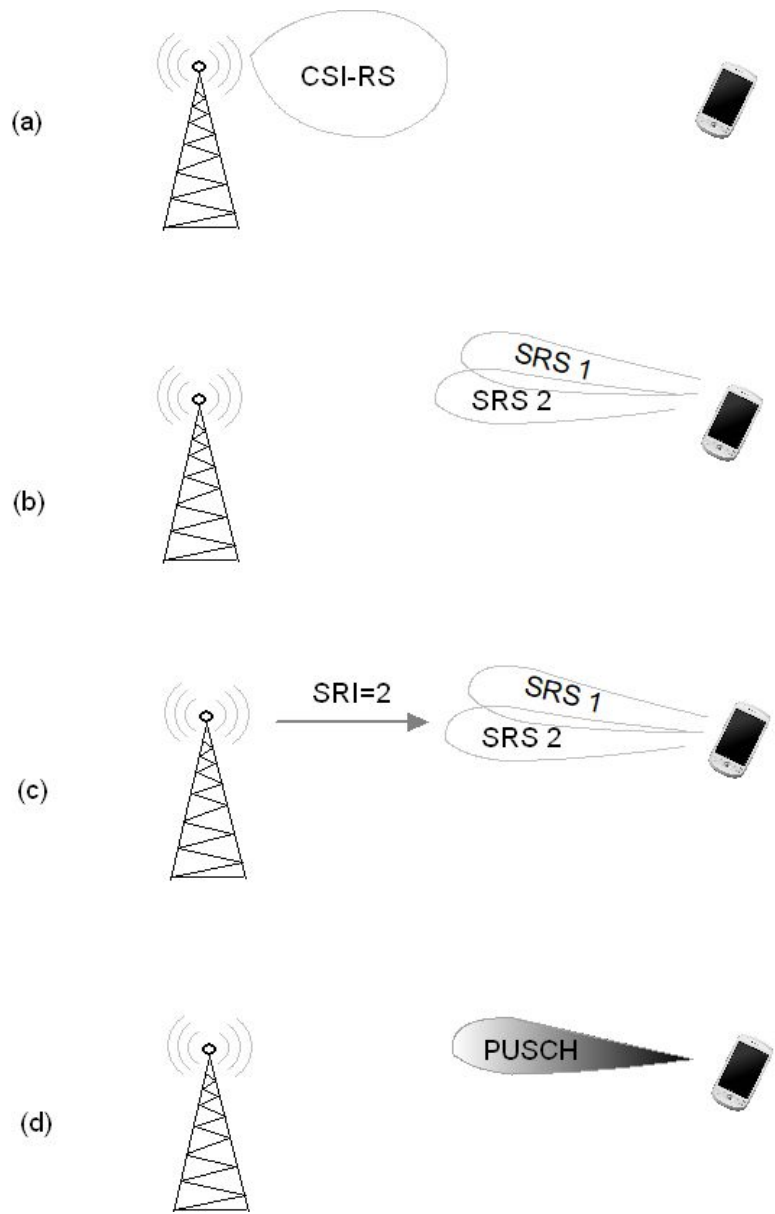

**Figure 7.** Non codebook-based UL transmission in NR. (a) The BS transmits CSI-RS to allow the UE estimate the DL channel; (b) the UE transmits up to 4 beamformed SRS; (c) the BS indicates a subset of the configured SRS using the SRI; (d) the UE carries PUSCH transmission using a reduced precoder.

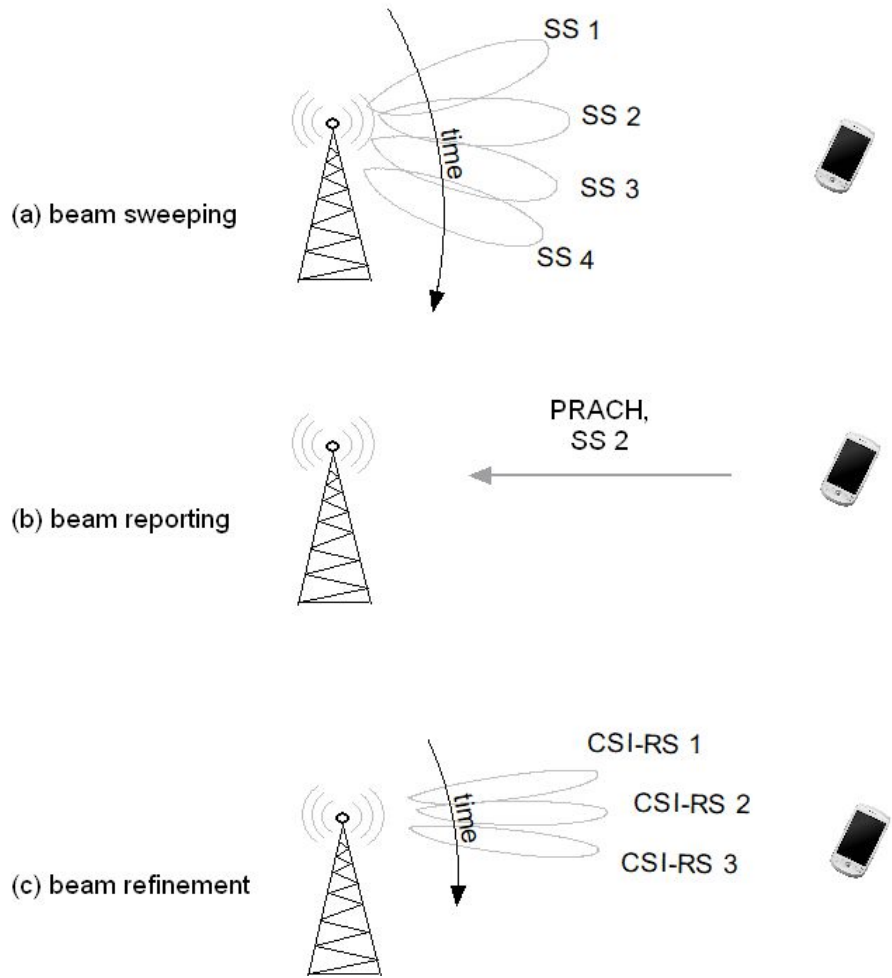

**Figure 8.** Beam management procedures. (a) Beam sweeping: the BS transmits up to 64 relatively wide beams within each SS block. (b) Beam reporting: the UE performs measurements on the SS beams and reports the best beam using the PRACH channel associated to the beam. (c) Beam refinement: the BS transmits a set of beamformed CSI-RS in a narrow angular sector centered around the wide SS beam.

## Tables

| | |
|---|---|
| TS 38.101-1 | UE radio transmission and reception; Part 1: Range 1 Standalone |
| TS 38.101-2 | UE radio transmission and reception; Part 2: Range 2 Standalone |
| TS 38.101-3 | UE radio transmission and reception; Part 3: Range 1 and Range 2 Interworking operation with other radios |
| TS 38.101-4 | UE radio transmission and reception; Part 4: Performance requirements |
| TS 38.104 | BS radio transmission and reception |
| TS 38.201 | Physical layer; General description |
| TS 38.202 | Services provided by the physical layer |
| TS 38.211 | Physical channels and modulation |
| TS 38.212 | Multiplexing and channel coding |
| TS 38.213 | Physical layer procedures for control |
| TS 38.214 | Physical layer procedures for data |
| TS 38.215 | Physical layer measurements |

**Table 1.** Main 3GPP 5G specifications related to the physical layer

| *Release* | *Downlink* | *Uplink* |
|---|---|---|
| Rel-8 (LTE) | transmit diversity, SM with CBP (4 layers) | N/A |
| Rel-10/13 (LTE-Advanced, LTE-Advanced Pro) | transmit diversity, SM with CBP/NCBP (8 layers) | SM with CBP (4 layers) |

**Table 2.** Main features of multiantenna techniques for the major LTE releases.

Related Articles (See Also)

| Article ID |
| --- |
| 5GRef001 |
| 5GRef003 |
| 5GRef008 |
| 5GRef021 |
| 5GRef023 |
| 5GRef024 |